\documentclass[11pt,nofootinbib,superscriptaddress,showkeys]{revtex4-1}%
\usepackage{dsfont}
\usepackage{amsmath}
\usepackage{amsfonts}
\usepackage{amssymb}
\usepackage{graphicx}%
\usepackage{subfig}
\usepackage{braket}
\usepackage{float}
\usepackage{pslatex}
\usepackage{slashed}
\usepackage{hyperref}
\usepackage{color}
\usepackage{tensor}
\usepackage[utf8]{inputenc}
\usepackage[T1]{fontenc}
\definecolor{darkgreen}{rgb}{0,0.35,0}

\newcommand{\p}{\partial}

\newcommand{\dd}{\ensuremath{\mathrm{d}}}

\newcommand{\be}{\begin{equation}}
\newcommand{\ee}{\end{equation}}
\newcommand{\bea}{\begin{eqnarray}}
\newcommand{\eea}{\end{eqnarray}}

\newcommand{\kulak}{KU Leuven Campus Kortrijk -- Kulak, Department of Physics, Etienne Sabbelaan 53 bus 7657, 8500 Kortrijk, Belgium}
\newcommand{\ughent}{Ghent University, Department of Physics and Astronomy, Krijgslaan 281-S9, 9000 Gent, Belgium}
\newcommand{\unesp}{Instituto de F\'{i}sica Te\'{o}rica, Universidade Estadual Paulista, Rua Dr.~Bento Teobaldo Ferraz, 271 - Bloco II, 01140-070 S\~{a}o Paulo, SP, Brazil}
\newcommand{\ucharles}{Faculty of Mathematics and Physics, Charles University, V Hole\v{s}ovi\v{c}k\'ach 2, 18000 Prague 8, Czech Republic}

\begin{document}

\title{Comments on the Chern--Simons photon term in the QED description of graphene}

\author{David Dudal}\email{david.dudal@kuleuven.be}\affiliation{\kulak}\affiliation{\ughent}
\author{Ana J\'{u}lia Mizher}\email{ana.mizher@kuleuven.be}\affiliation{\kulak}\affiliation{\unesp}
\author{Pablo Pais}\email{pablo.pais@kuleuven.be}\affiliation{\kulak}\affiliation{\ucharles}

\begin{abstract}
\noindent We revisit the Coleman--Hill theorem in the context of reduced planar QED. Using the global U(1) Ward identity for this non-local but still gauge invariant theory, we can confirm that the topological piece of the photon self-energy at zero momentum  does not receive further quantum corrections apart from the potential one-loop contribution, even when considering the Lorentz non-invariant case due to the Fermi velocity $v_F<c$. This is of relevance to probe possible time parity odd dynamics in a planar sheet of graphene which has an effective description in terms of $(2+1)$-dimensional planar reduced QED.
\end{abstract}


\maketitle


\section{Context and motivation}

Quantum Electrodynamics in $(2+1)$ dimensions (QED$_3$) has been widely used as a toy model for Quantum Chromodynamics (QCD). This is due to the fact that although being Abelian, QED$_3$ exhibits similar features as non-Abelian gauge theories, making it possible, for instance, to map and investigate chiral symmetry breaking and confinement into it \cite{Appelquist:1986qw,Appelquist:1986fd,Fischer:2004nq,Bashir:2008fk,Gusynin:2016som}. The similarity is reinforced by the fact that a non-Abelian gauge theory at high temperature suffers a dimensional reduction and, if coupled to $N_f$ fermion families, the non-Abelian interactions are suppressed by a factor of $N_f^{-1}$, so that in the large $N_f$ limit the theory can be considered approximately Abelian.

Recently, the emergence of the so-called Dirac and Weyl planar materials \cite{Wehling:2014cla}, converted QED$_3$ into a playground in which a potential link between high energy physics (including quantum fields in curved spacetimes) and condensed matter can emerge \cite{2006NatPh...2..620K,Vozmediano:2010zz,Cortijo:2011aa,Chernodub:2013kya,Li:2014bha,Iorio2012,Iorio2014}. Those are materials in which, due to the specific structure of their underlying lattice, the charge carriers present a relativistic-like behavior, being correctly described by a Dirac-like equation in some regimes. Particularly, the physical realization of graphene and other materials in two space dimensions, that are proved to contain a priori massless Dirac spinors, naturally yields the fermionic part of QED$_3$ \cite{Gusynin:2007ix,CastroNeto:2009zz} through the continuum limit of the tight-binding theory, usually applied to describe their conduction electrons, which in turn implies a direct connection to QCD, as discussed above.

Nevertheless, even though in these systems the fermions are constrained to remain in-plane and therefore are correctly described by a theory in $(2+1)$ dimensions, the gauge fields responsible for the interaction between these electrons are not subject to the same constraint. One of the most remarkable consequences of this fact is that the interaction between electrons remains the familiar $\sim1/r$ potential rather than the logarithmic one that would take place if the gauge fields were also restricted to the plane. Therefore, it is convenient and necessary to modify QED$_3$ in order to merge the desired features of the two sectors of the theory, starting with a general $(3+1)$ theory and dimensionally reducing it to a non-local effective $(2+1)$ theory. This procedure was followed within similar approaches in \cite{Marino:1992xi} with the so-called pseudo-QED (PQED), and posteriorly in \cite{Gorbar:2001qt}, receiving the name of reduced QED (RQED). In this work we follow the outline of RQED but both constructions are equivalent, and for interesting applications of PQED we refer to \cite{Alves:2013bna,Marino:2014oba,Marino:2015uda,Nascimento:2015ola}.

In the context of pure QED$_3$, the most general structure of the action allows for a term in the gauge sector that breaks time reversal ({\sf T}), namely the Chern--Simons (CS) term. Its presence gives a mass to the photon \cite{Deser:1981wh,Deser:1982vy} and, for this reason, it is also known as topological mass term (actually, in the Abelian case there is no real topology involved and the term ``topological'' is used for historical reasons based on its non-Abelian counterpart). This term is important in several contexts in condensed matter, for instance it leads naturally to the transverse conductivity observed from the Hall effect and it is crucial to model high $T_c$ superconductivity \cite{Chen:1989xs}. It was shown that radiative corrections coming from interaction terms can give a contribution for the topological photon mass up to one-loop. Remarkably, a theorem by Coleman and Hill \cite{Coleman:1985zi} demonstrates that, apart from one-loop, all corrections to the topological mass term vanish identically to all orders. This was done in general grounds, considering the photon interacting with any massive scalar, spinor or vector field with arbitrary gauge invariant interactions. The massive nature of the field excitations interacting with the photon is crucial here, to avoid the typical infrared subtleties in lower-dimensional field theories. In particular, the Coleman--Hill theorem does not hold in presence of massless degrees of freedom, as explicitly illustrated in e.g.~\cite{Spiridonov:1991ki}. Indeed, infrared singularities, typical for lower-dimensional field theories can disturb the argument.

Regarding the importance of RQED in the description of planar Dirac systems in condensed matter, precisely for those systems that allow for a direct analogy with QCD, it is important to verify if the Coleman--Hill theorem also holds for this theory, in particular when the Lorentz non-invariant version of RQED is considered. In this work we demonstrate that higher order radiative corrections are exactly vanishing in RQED, in the same way as for QED$_3$, meaning that the topological photon parameter arises at one-loop, or does not arise at all. In section \ref{sec:RQED} we discuss briefly how the tight-binding model yields QED$_3$ in the continuum limit and present the general features of RQED, including its gauge invariance and freedom of gauge choice, before and after the reduction. In particular we discuss possible mass terms for the fermions that are important if we want to apply our theory directly to graphene. The role of electromagnetic background fields in the radiative corrections, important in manipulations to study transport phenomena in materials, is also briefly highlighted, with explicit computations relegated to a future longer paper. In section \ref{sec:dudaltheorem} we prove in full detail, for the Lorentz invariant case, that corrections of order higher than one are null, then motivating our choice of mass terms, from both the (crucially different) two- as well as four-component spinor viewpoint, and finally summarizing the explicit one-loop computation in the absence of background fields. Section \ref{sec:dudaltheorem2} is devoted to the generalization of the argument to the Lorentz non-invariant case. In section \ref{sec:outlook} we present our final remarks.

\section{Setting the Stage: Planar systems and RQED}
\label{sec:RQED}

In this section we briefly review how the continuum limit of the tight-binding model describing graphene can be associated to QED in a lower dimension and why in this case it is interesting to work with a modified version of this theory, known as reduced QED. We discuss in some more detail the gauge invariance of this theory before and after the reduction is carried on, specially concerning the gauge fixing term, something not so well covered in other papers. Particular attention is paid to the role of fermion masses and how, in the continuum limit, different structures can result in equivalent mass terms, an issue that usually, although known \cite{Hands:2015qha}, is undervalued in the literature. Finally, we deduce the photon propagator for RQED taking into account a Chern--Simons term and we discuss the role of its coefficient, the $\theta$ parameter. We compare it to the standard QED$_3$, where $\theta$ is responsible for generating a photon mass and show that in RQED, although it also appears explicitly in the propagator, it differs dimensionally from a mass parameter, i.e.~the photon does remain massless for RQED.

As a starting point, we briefly present first the very basics of graphene from a point of view that is convenient for a quantum field theoretical approach. Many excellent reviews are available on this subject, as for instance \cite{Gusynin:2007ix,CastroNeto:2009zz} and references therein. Graphene, constituted by a single sheet of carbon atoms tightly packed into a two-dimensional honeycomb lattice, can be regarded in terms of two periodic sublattices $L_{A}$ and $L_{B}$. Here, we follow the convention of \cite{Gusynin:2007ix} (for an alternative convention see for instance \cite{Jackiw:2007rr}) and define the primitive two-dimensional vectors $\vec{a}_{i}$ for sublattice $L_{A}$ and $\vec{b}_{i}$ for the reciprocal sublattice, as ${\vec a}_1=a(1/2,\sqrt{3}/2)$, ${\vec a}_2=a(1/2,-\sqrt{3}/2)$ and ${\vec b}_1=\frac{2\pi}{a}(1/2,\sqrt{3}/2)$, ${\vec b}_2=\frac{2\pi}{a}(1/2,-\sqrt{3}/2)$, where $a$ is the sublattice spacing. It is also convenient to introduce the three near-neighbor vectors $\vec{s_{i}}$,
\begin{eqnarray}
  \vec{s}_{1}=a(0,1/\sqrt{3})\;, \quad\vec{s}_{2}=a(1/2,-{\sqrt{3}}/{6})\;,\quad  \vec{s}_{3}=a(-1/2,-{\sqrt{3}}/{6})\;,\end{eqnarray}
where $\ell=\frac{a}{\sqrt{3}}$ is the minimal lattice length.

The inner orbitals are strongly bonded to their respective carbon atom while the $\pi$ orbitals present a weak overlap. The electrons presented in these orbitals are called $\pi$ electrons. Following the usual tight-binding approach, only the interaction of each charge carrier with the nearest neighbors of $\pi$ electrons is considered. The Hamiltonian is written as
\begin{eqnarray}\label{Hamiltonian_pi_electrons}
\mathcal{H}&=&-t\sum_{\vec{r}\in L_{A}}\sum\limits_{i=1}^{3}\left(a^{\dagger}(\vec{r})b(\vec{r}+\vec{s_{i}})+b^{\dagger}(\vec{r}+\vec{s_{i}})a(\vec{r})\right)\;,
\end{eqnarray}
where the first sum is only along sublattice $L_{A}$, $t$ is the nearest-neighbor hopping energy and $a,a^{\dagger}(b,b^{\dagger})$ are the anticommuting ladder operators in the sublattice $L_{A}(L_{B})$. Applying a Fourier transformation it is straightforward to compute the energy-momentum dispersion relation \cite{CastroNeto:2009zz,Gusynin:2007ix}:
\bea
E(k_x,k_y)=\pm t \sqrt{3+2\cos\left({\sqrt{3}k_y a}\right)+4\cos\left({\frac{\sqrt{3}}{2}k_y a}\right)\cos\left({\frac{3}{2}k_x a}\right)}\;.
\eea
The valence and conduction band, generated by the opposite signs in the dispersion relation, touch in six points (Dirac points), of which only two are inequivalent. Here we choose them to be ${\vec K}_\pm = \pm 2\pi/a(2/3,0)$. Expanding the expression above around these zero energy points one can verify that the dispersion relation for each one of them is linear, $E_\pm ({\vec p})=\pm \hbar v_F|{\vec p}|$. Here, the Fermi velocity is determined by $v_F=\frac{3}{2}t\ell=\frac{\sqrt 3}{2}at\approx \frac{c}{300}$. It was shown \cite{Novoselov:2005kj} that the annihilation  operators $a$ and $b$ can be accommodated in a spinor field when we expand around the above Dirac points and, therefore, it can be seen as relativistic-like fermion that obeys a Dirac-like equation. In resume, the continuum limit of the nearest neighbors approach in a tight-binding model applied to a pure hexagonal sublattice with two intertwined triangular sublattices yields a massless version of the fermion sector of QED$_3$.

Following this approach and working with $\eta_{\mu\nu}=\mbox{diag}(-1,1,1)$, the action of the system reads:
\begin{equation}\label{action_pi_electrons}
\mathcal{S}_f=\int \dd^3x\left[\bar{\psi}\left[\gamma^0 (i\partial _0 -iv_F\vec{\gamma}\cdot \vec{\nabla} \right]\psi\right]\,,
\end{equation}
where here only the first two spatial gamma matrices $\vec{\gamma}$ enter. Here we show explicitly the Fermi velocity $v_F\leq1$ expressed in units of $c=1$. This is because later we shall deal with extra fields besides the fermion description of the $\pi$ electrons. If only these fermions were taken into account, we could take a simpler action in a Minkowskian space with a velocity $v_{F}$ instead of $c$ \cite{Chamon2007,Iorio2012,Iorio2014,Iorio2015}. In what follows, we will first focus on the $v_F=1$ limiting case, i.e.~the standard Lorentz invariant Dirac action. In Section \ref{sec:dudaltheorem2}, we will generalize the construction to the $v_F<1$ case.

Interactions with external sources or alterations on the underlying lattice, for instance using a substrate or doping, could produce a gap between the bands. This can be represented at the level of the action by a specific Dirac mass term, $m{\bf \bar{\psi}}{\bf \psi}$, or interaction terms involving the matter current. Let us refer to \cite{Gusynin:2007ix,Semenoff:1984dq,Ryu} for such possibilities and classification of the mass terms. Interaction terms that are bilinear in the fermion field will change the basic symmetries of the action, depending on their particular gamma matrices structure. In this paper we work in the chiral basis, where the gamma matrices and the fifth gamma matrix are given by:
\begin{eqnarray}
\gamma^0=\left(
\begin{array}{cc}
0 & I_2 \\
I_2 & 0
\end{array} \right),
\quad
\gamma^i=\left(\begin{array}{cc}
0 & \sigma^i \\
-\sigma^i & 0
\end{array}\right),
\quad
\gamma^5=\left(\begin{array}{cc}
-I_2 & 0 \\
0 & I_2
\end{array} \right)\,,
\qquad i=1,2,3\,,
\end{eqnarray}
with $I_{2}$ is the $2\times2$ identity and $\sigma^{i}$ are the standard Pauli matrices.
Among the several possibilities of interaction, one can observe that certain terms are completely equivalent to the Dirac mass term as they correspond to a change in the variables in the path integral. Since there is no axial anomaly in $(2+1)$ dimensions the result must describe the same physics. This is the case for the (anti-Hermitian) mass terms $m{\bf \bar{\psi}}\gamma^3{\bf \psi}$ and $im{\bf \bar{\psi}}\gamma^5{\bf \psi}$, that can be reached from the standard Dirac mass term by performing the following unitary transformations in the fermion fields \cite{Hands:2015qha}, respectively:
\begin{subequations}
\begin{eqnarray}
&&\psi \to e^{i\beta \gamma^5} \psi\ \ \ \  ;\ \ \ \bar{\psi} \to \bar{\psi} e^{i\beta \gamma^5}\,.\label{sym1}\\
&&\psi \to e^{\alpha \gamma^3} \psi\ \ \ \  ;\ \ \ \bar{\psi} \to \bar{\psi} e^{\alpha \gamma^3}\,,\label{sym2}
\end{eqnarray}
\end{subequations}
with appropriate choices of the ``angles'' $\alpha$ and $\beta$. In case of massless fermions, \eqref{sym1} and \eqref{sym2} both constitute symmetries of the theory and are part of a larger U(2) invariance, see \cite{Gusynin:2007ix}.

We remark that this is a feature of the continuum limit and discretization can bring differences between those terms. For example, the tight binding lattice models that would induce the three masses are different \cite{Semenoff:1984dq,Ryu}, but they share their continuum limit. Notice also that all these masses correspond to a {\sf T}-even sector \cite{Gusynin:2007ix}, where we refer to {\sf T}-even or {\sf T}-odd in the four-component spinor language. In the two-component description the symmetry behavior of the fermion mass terms can be different, see \cite{Jackiw:1980kv,Dunne:1998qy}.

Considering these variations of the Dirac mass in the continuum, it is particularly useful to go with $m{\bf \bar{\psi}}\gamma^3{\bf \psi}$ when working with a four-component representation of the fermion field, since in this way it is possible to decompose and rewrite the action in terms of two decoupled two-component spinors. This point will be discussed in more detail below in Section \ref{component}. The subtle differences between both formulations can also be appreciated from \cite{Jackiw:1980kv}.

Besides the variants of the Dirac mass, one other specific mass term is particularly important, the Haldane mass $m_o\gamma^3\gamma^5$ \cite{Haldane:1988zza}. This one is totally independent of the masses previously discussed, as it corresponds to a {\sf T}-odd bilinear term. The special interest in it relies on the fact that in pure QED$_3$ it can be directly related to the CS term.

The gauge sector of pure QED$_3$ is described by
\begin{equation*}
\mathcal{S}_{\text{QED}_3} = \int \dd^3x\left[-\frac{1}{4}F_{\mu\nu}F^{\mu\nu} +\frac{1}{2\xi}(\partial\cdot A)^2 - \frac{\theta}{2}\epsilon^{\mu\nu\rho}A_\mu \partial_\nu A_\rho\right]\,,
\end{equation*}
where the first is the usual Maxwell term, the second is a linear gauge fixing term and, the last one is the CS term.
On one hand, the one-loop radiative corrections from a fermion with Haldane mass generates a {\sf T}-odd piece in the photon polarization tensor \cite{Delbourgo:1992sh,Delbourgo:1994ws}, which can be translated  into the presence of the CS term in the gauge sector of the action. The Coleman--Hill theorem \cite{Coleman:1985zi} guarantees that no higher order corrections are allowed, so the connection of the two terms is clearly pictured. On the other hand, the presence of a CS term generates dynamically a Haldane mass for the fermions \cite{Bashir:2008ej} already at one-loop as well.

As discussed before, in order to correctly describe electrons confined to a plane but whose interaction is the usual Coulomb interaction, it is necessary to consider the gauge fields living in the three-dimensional spatial bulk rather than in the two-dimensional spatial plane. To obtain a consistent theory combining the suitable conditions for fermions and gauge fields, the authors in \cite{Marino:1992xi,Gorbar:2001qt} start with the gauge theory in four dimensions and integrate out the gauge field. Being deliberately brief, we consider standard QED$_4$ (without a Chern-Simons term) written as
\begin{eqnarray}\label{qed}
  \mathcal{S}_{\text{QED}_4} &=& \int \dd^4x \left[-\frac{1}{4}F_{\mu\nu}F^{\mu\nu}+\frac{1}{2\xi}(\partial\cdot A)^2+j_\mu A^\mu\right]\,.
\end{eqnarray}
The Dirac matter currents are supposed to be
\begin{eqnarray}
  j^{\mu} &=& \left\{\begin{array}{cc}
              i\bar \psi \gamma^\mu \psi \delta(x_3) &  ~\text{for}~\mu=0,1,2\,, \\
              0& ~\text{for}~\mu=3\,,
            \end{array}\right.\,,
\end{eqnarray}
with the fermion fields only dependent on $(x_0,x_1,x_2)$. This formally expresses the fact that the fermion dynamics is restricted to happen in the $(x_1,x_2)$-plane, i.e.~the planar graphene sheet. The current is conserved, $\p_\mu j^\mu=0$. The easiest way to proceed is to Wick rotate to Euclidean space and to Fourier transform (denoted by the~~$\hat{}$-notation throughout the remainder of the text) in order to integrate out the four-dimensional gauge field, leading to
\begin{eqnarray}\label{tussenin1}
  \mathcal{S}_{eff} &=& \int \dd^4p \left[{\hat j}^{\mu} {\hat D}_{\mu\nu}^T(\vec{p},p_3)  {\hat j}^{\nu}\right]\,,
\end{eqnarray}
where $\vec{p}=(p_0,p_1,p_2)$. $\hat D_{\mu\nu}^T(\vec{p},p_3)=\left(\delta_{\mu\nu}-\frac{p_\mu p_\nu}{(\vec{p}^2+p_3^2)}\right)\frac{1}{(\vec{p}^2+p_3^2)}$ is the (gauge independent) transverse projection of the free photon propagator, which appears due to the conserved fermion current.  As the Fourier-transformed currents will not depend on $p_3$, we can integrate out the latter, leading to
\begin{eqnarray}\label{tussenin2}
  \mathcal{S}_{eff} &=& \int \dd^3p \left[{\hat j}^{\mu} \hat{\mathcal{D}}_{\mu\nu}^T(\vec{p}) {\hat j}^{\nu}\right]\,.
\end{eqnarray}
The indices $\mu,\nu$ are from now on restricted to $x_0,x_1,x_2$ and we can forget about the $\delta(x_3)$ in the definition of the current $j_\mu$. Furthermore, we set
\begin{eqnarray}\label{tussenin3}
  \hat {\mathcal{D}}_{\mu\nu}^T(\vec{p})=\left(\delta_{\mu\nu}-\frac{p_\mu p_\nu}{p^2}\right)\frac{1}{2p}\,,\qquad p=\sqrt{\vec{p}^2}\,.
\end{eqnarray}
It is worth underlining that in passing from \eqref{tussenin1} to \eqref{tussenin2}, an irrelevant longitudinal term appearing in $\hat j_\mu\ldots\hat j_\nu$ has been dropped from \eqref{tussenin2}. It is then easily recognized that effective action \eqref{tussenin2} can be equivalently reformulated in terms of an Euclidean non-local gauge invariant three-dimensional theory, with gauge fixed action
\begin{eqnarray}\label{rqed}
\mathcal{S}_{\text{RQED}_3}=\int \dd^3 x \left[ \frac{1}{2}   F^{\mu\nu} \frac{1}{\sqrt{-\partial^2}} F_{\mu\nu}+\bar{\psi} (i\slashed{D} ) \psi + \frac{1}{2\zeta}(\partial\cdot A)^2 \right]\,,
\end{eqnarray}
after the introduction of a \emph{new} and now three-dimensional Abelian gauge field that, with a slight abuse of notation, we have again called $A_\mu$.  We have also added dynamics for the fermions, for the moment still without mass. The physical content of the theory will anyhow be gauge invariant and thus independent of the chosen gauge so that the gauge fixing term before or after the reduction does not need to be the same. We have opted here for a simple linear gauge fixing rather than the involved reduced non-local gauge fixing term kept in \cite{Marino:1992xi,Gorbar:2001qt}. The gauge parameter $\zeta$ here also carries a dimension, unlike $\xi$ in \eqref{qed}. The renormalization properties of RQED~$\equiv$ RQED$_{3}$ were discussed in \cite{Teber:2012de,Kotikov:2013eha}. It should be noted that \eqref{rqed} generates already at tree level a branch cut in the complex momentum plane in the photon propagator, with branch point at $p^2=0$. It is exactly the presence of the $1/\sqrt{-\partial^2}$ in the kinetic gauge term that also allows to keep the electromagnetic coupling constant $e$ to remain dimensionless, even in a (reduced) three-dimensional space-time. Indeed, the new gauge field $A_\mu$ still has mass dimension $1$, while for standard QED$_3$ that mass dimension would amount to $1/2$. The non-local operator $\sqrt{-\partial^{2}}^{-1}$ is to be understood   via its three-dimensional Fourier (momentum) space representation \cite{Marino:1992xi}
\begin{equation}\label{non-local-operator}
\frac{1}{\sqrt{-\partial^{2}}}(\vec{x}-\vec{x}') = \int \frac{d^{3}k}{2\pi^{3}} \frac{e^{i\vec{k}\cdot(\vec{x}-\vec{x}')}}{k}\,,\qquad k=\sqrt{\vec{k}^2} \;.
\end{equation}
If we add an Euclidean CS term, $i\theta\int \dd^3x~\epsilon_{\mu\nu\rho}A_{\mu}\partial_{\nu}A_{\rho}$ to the action in \eqref{rqed}, we can deduce the tree level photon propagator for a reduced Maxwell-CS theory, namely
\begin{equation}\label{photon_prop_tree}
\hat D_{\mu\nu}(\vec{p})=\frac{1}{2p}\frac{1}{(1+\theta^2)}\left(\delta_{\mu\nu}-\frac{p_\mu p_\nu}{p^2}\right)-\frac{1}{2p^2}\frac{\theta}{(1+\theta^2)}\epsilon_{\mu\nu\rho}p^{\rho}+\frac{\zeta}{p^2}\frac{p_\mu p_\nu}{p^2}
 \;.
\end{equation}
From the CS term, we can infer that $\theta$ here is actually a dimensionless parameter, so unlike in standard QED$_3$, it does not provide the theory with a ``topological photon mass''. This is consistent with the observation that RQED is scale invariant up to at least two-loops, i.e.~the beta function of the electromagnetic coupling vanishes\cite{Teber:2012de,Teber:2018goo}. On the other hand, $\theta\neq0$ does influence the photon propagator, not only by the presence of a {\sf T}-odd contribution, but also by a normalization of the photon propagator. Intuitively, this corresponds to a down-scaling of the strength of the photon propagator, an effect not unlike increasing the mass of the exchanged particle.

\section{One-loop exactness of topological photon term in reduced planar QED: Lorentz invariant case}
\label{sec:dudaltheorem}

Our aim is now to prove that there will be no $\sf T$-odd contributions to the gauge sector, i.e.~the CS term, coming from radiative corrections beyond one-loop. In principle, for the sake of physical interest, we could also try to add a generic electromagnetic background field to the action---be it to QED$_4$ or RQED$_3$ ---via the gauge principle of minimal coupling with the fermion fields. Background fields must be treated classically and, in the same way as the gauge sector, they must be defined in four dimensions. For possible interesting physics involving background fields see for instance \cite{Mizher:2013kza,Mizher:2016mfq,Mizher:2018dtf}, including in-plane fields as also considered in \cite{Aleiner:2007va}. For example, minimal coupling means we replace in \eqref{rqed} the covariant derivative as follows
\begin{eqnarray}\label{mincoup}
i\slashed{D}\to i\slashed{D} + i \bar A_0 \gamma^0 + i\bar A_3\gamma^3\,,
\end{eqnarray}
where the barred gauge fields are classical in nature. $\bar A_0$ can describe a potential (electric field $\vec{E}$) applied in or orthogonal to the graphene sheet, while $\bar A_3$ can be used to couple an in-plane magnetic field $\vec{B}\parallel \vec{e}_1$ . We remark here that the fields $A_{\mu}$ can be considered as the quantum fluctuations around these classical background fields $\bar{A}_{\mu}$. Taking the non-relativistic limit of the corresponding Dirac equation, the latter coupling will provide the necessary magnetic field-magnetic moment coupling relevant for the Zeeman term, considered in \cite{Aleiner:2007va}. It is important to realize that although graphene is a sheet and the fermions will have no \emph{classical} dynamics outside of the plane due to an in-plane magnetic field, there is still the option for further \emph{quantum} effects in the plane. Unfortunately, the tensorial basis elements relevant for the construction of a transverse self-energy, which play an important r\^{o}le in the Coleman--Hill argument, become far more complicated in presence of background fields, mostly due to the increased number of allowed transverse tensors in Fourier (momentum) space. Moreover, non-constant background fields make the situation utterly difficult. In the light of this, we will ignore background fields from our analysis in the current paper and we will from now on work with
\begin{eqnarray}\label{rqed2}
\mathcal{S}_{\text{RQED}_3}=\int \dd^3 x \left[ \frac{1}{2}  F^{\mu\nu} \frac{1}{\sqrt{-\partial^2}} F_{\mu\nu}+\bar{\psi} (i\slashed{D} + m\gamma^3 +m_o\gamma^3\gamma^5) \psi + \frac{1}{2\zeta}(\partial\cdot A)^2 \right]\,.
\end{eqnarray}
Notice that a $\vec{E}\cdot\vec{B}$ would be another {\sf T}-odd scalar quantity, if present. In the absence of such fields we have allowed for the Haldane mass as another source of {\sf T}-odd physics.  As explained before, we opted for the $m\bar\psi\gamma^3\psi$-representation of the Dirac mass, although the following argument does not depend on which fermion masses are present, the actual numbers can however.

\subsection{All order proof based on Ward identity}\label{proof}

First, we will use the power of the global Ward identity associated to charge conservation to prove that \eqref{rqed2} will generate a CS term for the photon at one-loop order, or not at all. It is important that the fermions are massive of some sort to avoid spurious infrared singularities, so we can hereafter safely consider zero momentum expansions. Such approach was suggested in \cite{Shacham:2013bma} for standard QED$_3$ whilst avoiding the combinatorial elements of the original proof of \cite{Coleman:1985zi}. We will follow as much as possible the analysis of \cite{Shacham:2013bma}, paying attention to some differences where necessary.

We decompose in Fourier space the three-dimensional photon $1PI$ propagator (self-energy) in its most general form in a linear covariant gauge that is compatible with all Ward (Slavnov-Taylor) identities,
\begin{eqnarray}\label{ch}
  \hat{\Pi}_{\mu\nu}(\vec{p}) &=& \braket{\hat A_\mu(\vec{p}) \hat A_\nu(-\vec{p})}^{1PI}= \left(\delta_{\mu\nu}-\frac{p_\mu p_\nu}{p^2}\right)\Pi(p^2) + \epsilon_{\mu\nu\rho}p^{\rho} \vartheta(p^2)\,. \end{eqnarray}
Although it is well known that the photon self-energy is transverse, let us shortly review the underlying argument, as we will also need it later on when we turn to the Lorentz non-invariant case (without change actually). It is most easily derived by replacing the action \eqref{rqed2} by its equivalent version
\begin{eqnarray}\label{rqedbrst}
\mathcal{S}_{\text{RQED}_3}=\int \dd^3 x \left[ \frac{1}{2}   F^{\mu\nu} \frac{1}{\sqrt{-\partial^2}} F_{\mu\nu}+\bar{\psi} (i\slashed{D}+i A_s \gamma^s +m\gamma^3 +m_o\gamma^3\gamma^5) \psi + b (\partial\cdot A) - \frac{\zeta}{2}b^2+\bar c \p^2 c\right]\,,
\end{eqnarray}
including the multiplier $b$ and Faddeev--Popov (anti-)ghost $\bar c, c$. Then the action \eqref{rqedbrst} enjoys a manifest BRST invariance, generated by
\begin{eqnarray}\label{brst2}
s A_\mu= -\partial_{\mu} c\,,\qquad  s\bar c= b\,,\qquad s c=0\,,\qquad s b=0\,,\qquad s\psi= -i e c\psi\,,\qquad s\bar \psi=-ie \bar\psi c\,,\qquad s^2=0\,.
\end{eqnarray}
We can define the composite operators $s\psi$ and $s\bar\psi$ at the quantum level by means of appropriate external sources coupling them to the theory,
\begin{eqnarray}\label{rqedbrstbis}
\Sigma=\mathcal{S}_{\text{RQED}_3}+\int \dd^3 x \left[\bar{\mathcal{J}}s\psi-s\bar\psi\mathcal{J}\right]\,.
\end{eqnarray}
At the functional level, the BRST invariance is encoded in
\begin{eqnarray}\label{brst3}
  \int \dd^3x\left[-\p_\mu c\frac{\delta \Sigma}{\delta A_\mu }+b\frac{\delta \Sigma}{\delta \bar c }+\frac{\delta \Sigma}{\delta \bar {\mathcal{J}} }\frac{\delta \Sigma}{\delta \psi }-\frac{\delta \Sigma}{\delta  \mathcal{J} }\frac{\delta \Sigma}{\delta \bar\psi }\right]=0\,,
\end{eqnarray}
which becomes the Slavnov--Taylor identity at the quantum level,
\begin{eqnarray}\label{brst4}
\int \dd^3x\left[-\p_\mu c\frac{\delta \Gamma}{\delta A_\mu }+b\frac{\delta \Gamma}{\delta \bar c }+\frac{\delta \Gamma}{\delta \bar {\mathcal{J}} }\frac{\delta \Gamma}{\delta \psi }-\frac{\delta \Gamma}{\delta  \mathcal{J} }\frac{\delta \Gamma}{\delta \bar\psi }\right]=0\,.
\end{eqnarray}
Here, $\Gamma$ is the quantum effective action, viz.~the generating functional for the $1PI$ correlation functions. We have also suppressed the space time variable $x$ to avoid notational clutter. Acting with the test operator $\frac{\delta^2}{\delta c \delta A_\nu }$ on \eqref{brst4} and setting all external fields and sources to null, we obtain indeed the well-known transversality constraint
\begin{eqnarray}\label{brst5}
\partial_{\mu} \frac{\delta^2 \Gamma}{\delta A_\mu \delta A_\nu}\equiv  \partial_{\mu} \Pi^{\mu\nu}=0\,.
\end{eqnarray}
Returning to the decomposition \eqref{ch}, the Coleman--Hill theorem now states that $\lim_{p^2\to0}\vartheta(p^2)$ is solely determined by one-loop corrections.

To show this explicitly, we start from the path integral,
\bea\label{pi1}
\mathcal{I}&=&\int [\dd\bar\psi][\dd\psi][\dd A_\mu] e^{-\mathcal{S}_{\text{RQED}_3}}\,,
\eea
with $\mathcal{S}_{\text{RQED}_3}$ defined in \eqref{rqed2}. Then diagrammatically it is easily seen that at zero momentum, the graphs contributing to ${\hat\Pi}_{\mu\nu}(p^2)$ are corresponding to those of the $1PI$ current-current correlator with zero momentum flow. We shall hence focus attention on $\braket{j_\mu(x) j_\nu(y)}^{1PI}$ and show that at zero momentum, it is fully determined at one-loop order.

Classically, we can couple the current $j^{\mu}(x)$ to the action via an extra local source $\eta_\mu(x)$ by considering
\bea\label{loc1}
\Sigma'=\Sigma+\int \dd^3x~\eta_\mu j^{\mu}\,,
\eea
then
\bea\label{loc2}
\p_\mu\frac{\delta\Sigma'}{\delta\eta_\mu}= \bar\psi\frac{\delta\Sigma'}{\delta\bar\psi}+\frac{\delta\Sigma'}{\delta\psi}\psi
\eea
expresses that the current is conserved. This is nothing else than the Noether theorem in functional language. As before, we get the quantum Ward identity,
\bea\label{loc3}
\p_\mu\frac{\delta\Gamma}{\delta\eta_\mu}= \bar\psi\frac{\delta\Gamma}{\delta\bar\psi}+\frac{\delta\Gamma}{\delta\psi}\psi\,.
\eea
Here, we tacitly ignore the presence of the non-local operator $\frac{1}{\sqrt{-\p^2}}$, as strictly speaking, the quantum validity of Ward identities is only ensured in terms of local quantum field theory \cite{Piguet:1995er}. This being said, the non-locality can be reduced from  $\frac{1}{\sqrt{-\p^2}}$ to $\sqrt{-\p^2}$ by introducing an auxiliary anti-symmetric tensor field $B_{\mu\nu}$ that allows to replace $\int \dd^3 x \frac{1}{2}  F^{\mu\nu} \frac{1}{\sqrt{-\partial^2}} F_{\mu\nu}\to\int \dd^3 x \left[ \frac{1}{2} B^{\mu\nu} \sqrt{-\partial^2} B_{\mu\nu}+  B_{\mu\nu}F^{\mu\nu}\right]$. This replacement would not affect the rest of the proof in the section, but a complete localization remains impossible, unless by going back to a higher dimension of course, i.e.~the original starting point with its mixed-dimensional action. Though, it is also important to realize that the non-local term only affects the photon propagator, making it even softer in the infrared than for standard QED$_3$. As such, the infrared is safer than expected, while for the interaction terms (leading to the non-trivial Feynman diagrams), standard power counting will apply. This is also the reason we could present the current analysis, which makes clear that the RQED situation is, although much alike, not completely similar to its better known cousin QED$_3$, as treated in \cite{Shacham:2013bma}. Power counting renormalizability of RQED was discussed in \cite{Teber:2012de,Teber:2018goo,Herzog:2017xha}. This ensures that the tree level non-locality will not spread into the theory, in that sense that there is no need to introduce more and more interaction vertices into the theory to maintain renormalizability. For example, if higher powers of $\frac{1}{\sqrt{-\p^2}}$ were to be combined with higher powers of $F_{\mu\nu}$, still $d=3$ but higher order interaction vertices could appear radiatively. This is not the case for RQED. As $\dim[\eta_\mu]=1$, the quantum version of the action \eqref{loc1}, which should contain all possible integrated $d=3$ polynomials of fields and sources compatible with the Ward identity constraints, will not be deformed by terms containing $\eta_\mu^2$ or higher powers, as no such terms can be constructed. The latter type of terms, if present, are responsible for contact terms in correlation functions containing the (gauge invariant) current, see e.g.~\cite{Dudal:2008tg} for a similar observation. We will thus not need to worry about contact terms from the start, thereby evading the comment of \cite{Shacham:2013bma}.

Let us now denote with $\mathcal{V}_0\equiv -i\int \dd^3x {\bf \bar{\psi}} A_\mu \gamma^\mu{\bf \psi}$ the standard gauge-boson fermion vertex operator. Then we can infer from the Ward identity \eqref{loc3} that
\begin{eqnarray}\label{pi5}
  \partial^{\mu}\Braket{j_\mu(x) j_\nu(y)\mathcal{V}_0^n}^{1PI} &=& 0
\end{eqnarray}
by taking another functional derivative of \eqref{loc3} w.r.t.~$\eta_\nu(y)$, followed by $n\geq0$ derivatives w.r.t.~the coupling constant $e$ and setting all external sources and fields to zero at the end. The coupling $e$ acts here as the (constant) source defining by its derivatives the quantum insertion of the vertex operator $\mathcal{V}_0$.  Notice that each power of $\mathcal{V}_0$ is an integrated operator insertion, that is, one with zero momentum flow. Since \eqref{pi5} holds for any $n$ and since any expectation value of operators evaluated with the path integral partition function \eqref{pi1} can be succinctly rewritten as
\begin{eqnarray}\label{pi6}
\Braket{j_\mu(x) j_\nu(y)}_{\mathcal{S}_{\text{RQED}_3}}^{1PI}=\sum_{n\in\mathbb{N}}\braket{j_\mu(x) j_\nu(y)e^n\mathcal{V}_0^n}_{\text{quad}}^{1PI}\,,
\end{eqnarray}
where ``$\text{quad}$'' refers to the quadratic (free theory) approximation of $\mathcal{S}_{\text{RQED}_3}$, we can equally well write
\begin{eqnarray}\label{pi7}
\sum_{n\in\mathbb{N}}\partial^{\mu}\braket{j_\mu(x) j_\nu(y)e^n\mathcal{V}_0^n}_{\text{quad}}^{1PI}=0
\end{eqnarray}
instead of \eqref{pi5}.

For $n\geq 0$, each term in the expansion \eqref{pi6} can be expanded around zero momentum as
\begin{eqnarray}\label{pi8}
\braket{\hat j_\mu(p) \hat j_\nu(-p)\mathcal{\hat V}_0^n}_{\text{quad}}^{1PI}= a_n \delta_{\mu\nu} + b_n \epsilon_{\mu\nu\rho}p^{\rho}+\ldots
\end{eqnarray}
after Fourier transforming. There is no need for the transverse projector $P_{\mu\nu}(p)=\delta_{\mu\nu}-\frac{p_\mu p_\nu}{p^2}$ here, as for $p\to0$, this operator becomes proportional to $\delta_{\mu\nu}$.

Combining the constraint \eqref{pi7} with \eqref{pi8} then immediately gives $a_n=0$ for all $n\geq 0$. To control the $b_n$, we use a small trick. We replace
\begin{eqnarray}\label{pi9}
\braket{\hat j_\mu(p) \hat j_\nu(-p)\mathcal{\hat V}_0^n}_{\text{quad}}^{1PI}\to \lim_{k\to0}\braket{\hat j_\mu(p+k/2) \hat j_\nu(-p+k/2)\mathcal{\hat V}_0^{n-1}\mathcal{\hat V}_k}_{\text{quad}}^{1PI}\,,
\end{eqnarray}
i.e.~we let a small net momentum $k$ flow through one of the vertices, keeping total momentum conservation in mind of course. Strictly speaking from the viewpoint of renormalization, we should introduce here another local source to define the non-integrated quantum operator $-i{\bf \bar{\psi}} A_\mu \gamma^\mu{\bf \psi}$, thereby deforming again the original action \eqref{loc1}. However, since we are only interested in the zero momentum limit, i.e. integrated operator, we refrain from doing so. This means we must exclude the $n=0$ term as we need at least one vertex insertion. Due to the symmetry $(\mu,p)\leftrightarrow(\nu,-p)$ present in expression \eqref{pi9}, only the following expansion can hold at leading order in $(p,k)$,
\begin{eqnarray}\label{pi10}
\braket{\hat j_\mu(p+k/2) \hat j_\nu(-p+k/2)\mathcal{\hat V}_0^{n-1}\mathcal{\hat V}_k}_{\text{quad}}^{1PI}=A_n\delta_{\mu\nu}+B_n \epsilon_{\mu\nu\rho}p_\rho +\ldots\,.
\end{eqnarray}
Since $k$ does not appear in the foregoing expression, we actually have $B_n=b_n$ for $n\geq 1$ from the identification \eqref{pi9} together with the expansion \eqref{pi8}. This aforementioned symmetry is the crucial part to discard other possible momentum combinations in \eqref{pi10}, we assume that \cite{Shacham:2013bma} used the same observation, without having made it explicit though.

The Fourier version of the constraint\footnote{This condition also holds when the operators $\mathcal{\hat V}_k$ would not be integrated, this can be shown by coupling the operator $\bar\psi \slashed{A}\psi$ to the action $\Sigma$ with another local source and by manipulating the corresponding Ward identity.} \eqref{pi7} now reads
\begin{eqnarray}\label{pi11}
(p+k/2)^{\mu}\braket{\hat j_\mu(p+k/2) \hat j_\nu(-p+k/2)\mathcal{\hat V}_0^{n-1}\mathcal{\hat V}_k}_{\text{quad}}^{1PI}=0 \,.
\end{eqnarray}
Applying this to \eqref{pi10} leads, next to $A_n=0$, to $b_n=0$ for all $n\geq 1$.

Putting everything back together, we have actually shown that
\begin{eqnarray}\label{pi12}
\Braket{\hat j_\mu(p) \hat j_\nu(-p)}_{\mathcal{S}_{\text{RQED}_3}}^{1PI}=b_0\epsilon_{\mu\nu\rho}p_\rho + \mathcal{O}(p^2)\,,
\end{eqnarray}
which is nothing else than the Coleman--Hill theorem for RQED, as the corresponding zeroth order diagram contributing to \eqref{pi12} is equivalent to the one-loop photon self-energy correction.

\subsection{Four-component vs.~two-component spinors}
\label{component}

As we mentioned in Section \ref{sec:RQED}, there are several theoretical instances to create a mass gap in the Dirac regime of graphene $\pi$-electrons, even if experimentally it is still very difficult to open a mass gap in a controllable way \cite{CastroNeto:2009zz} (see \cite{Gusynin:2007ix,Ryu,ConchaSanchez:2013cp} for a detailed description of the different mass terms and their corresponding symmetry breaking). Here, we shall briefly survey how those mass terms reduce in the four- and two-component spinor description for these electrons. If we consider four-spinors in $(2+1)$ dimensions, the Lorentz generators are in a reducible $4\times4$ matrix representation \cite{Gusynin:2007ix,Hands:2015qha,Iorio2015}. We arrange the sublattice annihilation operators ($a$ and $b$) with their corresponding valley numbers (subscript $+$ and $-$ ) as
\begin{equation}\label{2-spinor}
\psi_{+}=\left(
       \begin{array}{c}
         a_{+} \\
         b_{+} \\
       \end{array}
     \right)\;,\; \psi_{-}=\left(
       \begin{array}{c}
         b_{-} \\
         a_{-} \\
       \end{array}
     \right)\;,
\end{equation}
in two-component representation, and as
\begin{equation}\label{4-spinor}
\psi=\left(
       \begin{array}{c}
         \psi_{+} \\
         \psi_{-} \\
       \end{array}
     \right)\;,
\end{equation}
in the case of a four-component representation.

As in the four-component description we have at our disposal two matrices which anti-commute with respect to the rest ($\gamma^{3}$ and $\gamma^{5}$), we have basically four kinds of masses. Notice that we do not consider the internal spin-$1/2$ nature of the $\pi$ electrons. Considering it, the number of mass terms increases considerably \cite{Ryu}. A standard mass term in four-component spinor language is of the form $m\overline{\psi}\psi=m\psi^{\dagger}\gamma^{0}\psi$, which breaks both symmetries \eqref{sym1} and \eqref{sym2}, but it does not break time reversal symmetry in the four-dimensional matrix representation. This term mixes the flavours $+$ and $-$,
\begin{equation*}
S_{\text{usual}}=-\int \dd^3xm\overline{\psi}\psi=-\int \dd^3x m\left(\psi^{\dagger}_{+}\psi_{-}+\psi^{\dagger}_{-}\psi_{+}\right)\;.
\end{equation*}
The mass terms considered in Section \ref{sec:RQED}, i.e., $im{\bf \bar{\psi}}\gamma^5{\bf \psi}$ and $m{\bf \bar{\psi}}\gamma^3{\bf \psi}$, break one of the extended symmetries, \eqref{sym1} and \eqref{sym2} respectively, but preserve time reversal symmetry in four-dimensional matrix representation. The first case is related to the Kekul\'e distortion \cite{Hou:2006qc}, while we can see that the second one allow us to rewrite the action in a two-component spinor decomposition as
\begin{equation}\label{action_gamma3}
S_{\gamma^{3}}=-\int \dd^3x m\overline{\psi}\gamma^{3}\psi=-\int \dd^3x m\left(\psi^{\dagger}_{+}\sigma^{3}\psi_{+}-\psi^{\dagger}_{-}\sigma^{3}\psi_{-}\right)\;.
\end{equation}
We will call this term the ``normal'' mass, as is the usual mass for a two-component spinor in $(2+1)$ dimensions with two different decoupled flavours $+$ and $-$.

The last possibility is the Haldane mass term \cite{Haldane:1988zza}, which does not break the symmetries \eqref{sym1} and \eqref{sym2}, but  does break time reversal symmetry \cite{Gusynin:2007ix}. This term also admits a decoupled two-component spinor decomposition,
 \begin{equation}\label{action_haldane}
S_{\text{Haldane}}=-\int \dd^3x m_{0}\overline{\psi}\gamma^{3}\gamma^{5}\psi=-\int \dd^3x m_{o}\left(\psi^{\dagger}_{+}\sigma^{3}\psi_{+}+\psi^{\dagger}_{-}\sigma^{3}\psi_{-}\right)\;.
\end{equation}
We can see that the mass terms \eqref{action_gamma3} and \eqref{action_haldane} have different relative sign for the two flavours $+$ and $-$.

The CS mass term can be generated by {\sf T}-odd fermion one-loop corrections. These corrections at zero momentum are of the form \cite{RedlichPRL,RedlichPRD,Delbourgo:1994ws,Dunne:1998qy}
\begin{equation}
\Gamma_{\mu\nu}^{\text{odd}}\sim \frac{m}{|m|}\epsilon_{\mu\nu\rho}p_{\rho}\;,
\end{equation}
implying that the term \eqref{action_gamma3} will give a net zero contribution for the CS photon mass, while \eqref{action_haldane} does contribute. More precisely, we will get at the level of the action a (exact) radiatively introduced {\sf T}-photon term
\begin{equation}\label{csdyn}
S_{\text{CS}}=\int \dd^3x \left(-i\frac{e^2}{4\pi}\frac{m_o}{|m_o|}\epsilon^{\mu\nu\rho}A_\mu \p_\nu A_\rho\right)
\end{equation}
when a Haldane term \eqref{action_haldane} is coupled to RQED. Here is a nice place to appreciate again the role of the dimensionless coupling in RQED. Indeed, in the case of QED$_3$ the $e^2$ in front of \eqref{csdyn} is what ``feeds'' the dynamical topological photon \emph{mass} $\theta$ thanks to $e^2$ having mass dimension 1, whereas now the dimensionless nature of $e^2$ gives a \emph{dimensionless parameter} $\theta$ in front of the CS term.

\section{One-loop exactness of topological photon term in reduced planar QED: Lorentz non-invariant case}
\label{sec:dudaltheorem2}
Having proven the Coleman--Hill theorem in the case of Lorentz-invariant RQED$_3$, let us now turn to the generalization in terms of the action (based on the one of \eqref{action_pi_electrons}, supplemented with the photon field and BRST invariant linear gauge fixing)
\begin{eqnarray}\label{rqedbrstlor1}
\mathcal{S}_{\text{RQED}_3}=\int \dd^3 x \left[ \frac{1}{2}   F^{\mu\nu} \frac{1}{\sqrt{-\partial^2}} F_{\mu\nu}+\bar{\psi} (i \gamma^0 (\p_0+eA_0)+i v_F \gamma^i(\p_i+eA_i)+ m\gamma^3 +m_o\gamma^3\gamma^5) \psi + b (\partial\cdot A) - \frac{\zeta}{2}b^2+\bar c \p^2 c\right]\,,\nonumber\\
\end{eqnarray}
i.e.~we take into account the Fermi velocity $v_F$. To avoid further notational clutter, we shall keep the notation $\mathcal{S}_{\text{RQED}_3}$ for the classical action, $\Sigma$ for the classical action supplemented with external sources and $\Gamma$ for the quantum effective action. We will be brief about the points that do not change, but go into more detail into the necessary significant adaptations. Some quantum aspects of a similar theory---for massless fermions and within the approximation of an instantaneous Coulomb interaction---were discussed in \cite{Son:2007ja}, including the renormalization of $v_F$ when away from the fixed point $v_F=1$ (also studied in \cite{Herbut2009}), the latter corresponding to the Lorentz invariant limit. Dyson--Schwinger equation based studies are presented in e.g.~\cite{Carrington:2016fsh,Carrington:2017hlc}.

Clearly, the action \eqref{rqedbrstlor1} is still BRST invariant w.r.t.~\eqref{brst2}, so the transversality constraint \eqref{brst5} holds, irrespective of $v_F$. As the Lorentz invariance is reduced to two-dimensional rotational invariance in the $(1,2)$-plane, the tensorial decompositions as used in Section \ref{sec:dudaltheorem} become a bit more elaborate. The self-energy can now be decomposed into
\begin{eqnarray}\label{chnieuw}
  \Pi_{\mu\nu}(\vec{p}) &=& \braket{A_\mu(\vec{p}) A_\nu(-\vec{p})}^{1PI}= P_{\mu\nu}^1\Pi_1(p_0,p_ip_i)+P_{\mu\nu}^2\Pi_2(p_0,p_ip_i) + \epsilon_{\mu\nu\rho}p_\rho \vartheta(p_0,p_i p_i)\,. \end{eqnarray}
We have introduced two transverse projectors, similar to as what is known from finite temperature field theory \cite{Kapusta:2006pm,Brandt:2000tf}, as the raison d'\^{e}tre for the relevance of these two tensors is the same: the $0$-direction is singled out as ``special''. To be more precise, we have, with $i,j\in\{1,2\}$
\begin{eqnarray}
  P_{\mu\nu}^{(1)} &=& \left\{\begin{array}{ccc}
                     0 & \quad & \mu=0 ~\text{or}~ \nu=0 \\
                     \delta_{ij}-\frac{p_ip_j}{p^2} & \quad & \text{otherwise}
                   \end{array}\right.\,,\qquad P_{\mu\nu}^{(2)} = \left(\delta_{\mu\nu}-\frac{p_\mu p_\nu}{p^2}\right)-P_{\mu\nu}^{(1)}\,.
\end{eqnarray}
The decomposition \eqref{chnieuw} is the most general one that is compatible with the symmetry $(\mu,p)\leftrightarrow(\nu,-p)$, the two-dimensional rotational invariance and the transversality constraint $p_\mu \Pi_{\mu\nu}=0$. The form factors can depend separately on $p_0$ and $p_ip_i$, as indicated.

The (conserved) fermion current is now given by
\begin{equation}\label{currentbis}
  j_\mu= \bar\psi i\gamma^0\psi\delta_{\mu0}+v_F\bar\psi i\gamma^i\psi\delta_{\mu i}\,,\qquad \partial^{\mu} j_\mu=0\,,
\end{equation}
while the photon-fermion vertex becomes $\mathcal{V}_0= -i\int \dd^3x \bar\psi A_0 \gamma^0 -iv_F\int \dd^3x \bar\psi A_i \gamma^i$. This vertex could be split into 2 vertices, but considering all powers of $\mathcal{V}_0$ in the series expansion will generate all necessary powers of its 2 substructures, so we can maintain a single vertex expression for simplicity.

The connection between the self-energy $\Pi_{\mu\nu}(\vec{p})$ and the current-current correlator remains valid, so it is still sufficient to control the low momentum expansion of $\braket{j_\mu j_\nu}^{1PI}$, which is rewritable as, using the same arguments as in eq.~\eqref{pi6}
\begin{eqnarray}\label{pi6bis}
\Braket{j_\mu j_\nu}_{\mathcal{S}_{\text{RQED}_3}}^{1PI}=\sum_{n\in\mathbb{N}}\braket{j_\mu j_\nu e^n\mathcal{V}_0^n}_{\text{quad}}^{1PI}
\end{eqnarray}
and subject to \eqref{pi7}, with the appropriate vertex replacement.

A word of caution is in place here. As the fermions are massive, just as before there will be no singular behavior around zero momentum, despite the lack of Lorentz invariance. This situation is in sharp contrast with finite temperature, where non-analytic behavior emerges at zero momentum, see for instance \cite{Brandt:2000tf,Das:1997gg}. In particular do the limits $p_0\to0$ and $p_i\to0$ not commute, as they correspond to different physics. Intuitively, a finite temperature medium opens extra reaction channels (particle absorption from the medium), leading to extra branch points, in particular at the momentum origin \cite{Das:1997gg}. This lies at the heart of the non-validity of the Coleman--Hill theorem at finite temperature, explicitly illustrated in \cite{Brandt:2000tf}. Our current setup is inherently different, as we have no thermal medium. As such, for each $n\geq0$, we can expand the terms in \eqref{pi6bis} as follows:
\begin{eqnarray}\label{pi8bis}
\braket{\hat j_\mu(p) \hat j_\nu(-p)\mathcal{\hat V}_0^n}_{\text{quad}}^{1PI}= a_n \delta_{\mu0}\delta_{\nu0} + a_n'\delta_{\mu i}\delta_{\nu i} + b_n \epsilon_{\mu\nu\rho}p_\rho+\ldots\,.
\end{eqnarray}
The two delta-terms correspond to the zero momentum limits of the aforementioned transverse projectors. The rest of the argument proceeds analogously as in Section \ref{sec:dudaltheorem}, eventually leading to $a_n=a_n'=0, \forall n\geq0$, next to $b_n=0, \forall n\geq1$. Said otherwise, the zero momentum limit of the topological term in the photon self-energy is exact at one-loop order, i.e.~the Coleman--Hill theorem applies to the theory described by the action \eqref{rqedbrstlor1}.

It is a nice exercise to compute $\vartheta(0)$ and see how it depends on $v_F$. We will follow \cite{Dunne:1998qy,Chaichian:1997ix} and consider first a single two-component (Euclidean) spinor with standard Dirac mass. We use \begin{eqnarray}\gamma^0=\left(
\begin{array}{cc}
-i & 0 \\
0 & i
\end{array} \right)\,,\qquad\gamma^1=\left(
\begin{array}{cc}
0 & i \\
i & 0
\end{array} \right)\,,\qquad\gamma^2=\left(
\begin{array}{cc}
0 & 1 \\
-1 & 0
\end{array} \right)\,.\end{eqnarray}
The one-loop photon self-energy is then given by
\begin{equation}\label{p1}
  \Pi_{\mu\nu}(\vec p)= e^2 \int \frac{\dd^3k}{(2\pi)^3}\text{Tr}\left[(\gamma^0\delta_{\mu0}+v_F\gamma^i\delta_{\mu0})S_F(\vec{p}+\vec{k})(\gamma^0\delta_{\nu0}+v_F\gamma^j\delta_{\nu j})S_F(\vec k)\right]\,,
\end{equation}
with a fermion propagator reading
\begin{equation}\label{p2}
S_F(\vec{p})=\frac{p_0\gamma^0+v_F p_i\gamma^i-m}{p_0^2+v_F^2p_i p_i}\,.
\end{equation}
To facilitate the computation, we first notice that we can introduce $\vec{P}=(p_0,v_F p_i)$, and doing the same for the integration momentum $\vec{K}=(k_0,v_F k_i)$, we get
\begin{equation}\label{p3}
  \hat \Pi_{\mu\nu}(\vec p)\equiv \tilde\Pi_{\mu\nu}(\vec P)= \frac{e^2}{v_F^2} \int \frac{\dd^3K}{(2\pi)^3}\text{Tr}\left[(\gamma^0\delta_{\mu0}+v_F\gamma^i\delta_{\mu0})\frac{\slashed{P}+\slashed{K}-m}{(P+K)^2+m^2}(\gamma^0\delta_{\nu0}+v_F\gamma^j\delta_{\nu j})\frac{\slashed{K}-m}{K^2+m^2}\right]\,.
\end{equation}
As we are only interested in the piece $\propto \epsilon_{\mu\nu\rho}p^{\rho}$, it is clear that the only relevant contributions to this odd piece can come from combining an odd number of $\gamma$-matrices, based on the property $\text{Tr}(\gamma^\mu \gamma^\nu \gamma^\rho)=-2\epsilon^{\mu\nu\rho}$. It is clear from the above expression \eqref{p3} that we will find for the integral, at leading order in $P$ and thus in $p$, the same result as if $v_F=1$, modulo the fact that each time a spatial index appears, an extra factor of $v_F$ is to be included, either coming from the $\gamma^i$-and/or $\gamma^j$-vertex, or from the spatial part of the $P_\rho$-factor multiplying $\epsilon_{\mu\nu\rho}$. Thanks to the $\epsilon$-symbol, we know that exactly two such spatial indices will appear in any case, so keeping into account the prefactors of the integral, we will ultimately find
\begin{equation}\label{p4}
  \hat\Pi_{\mu\nu}(\vec p)=\frac{e^2}{4\pi}\frac{m}{|m|}\epsilon_{\mu\nu\rho}p^{\rho}+\mathcal{O}(p^2)\,,
\end{equation}
i.e.~the topological photon term does not depend on the Fermi velocity. This result confirms the earlier finding of \cite{Fialkovsky:2016kio}, where a $v_F$-rescaling of the spatial $\gamma$-matrices was introduced to facilitate the one-loop computation of the self-energy.

Returning to the graphene case with four-component spinors, we are thus led to no dynamically generated CS-term in presence of a Dirac mass, while a Haldane mass leads to
\begin{equation}\label{csdynbis}
S_{\text{CS}}=\int \dd^3x \left(-i\frac{e^2}{4\pi}\frac{m_o}{|m_o|}\epsilon^{\mu\nu\rho}A_\mu \p_\nu A_\rho\right)
\end{equation}
if the underlying dynamics is governed by the action \eqref{rqedbrstlor1}.

\section{Outlook}
\label{sec:outlook}

We have shown that, in the framework of reduced QED in $(2+1)$ dimensions, the topological piece of the photon self-energy at zero momentum only receive quantum corrections up to one-loop. Using fundamental arguments based on the $U(1)$ Ward identity, we have proven that all the two- and higher-loop contributions are identically zero. In other words, besides holding for ordinary QED$_3$, the Coleman--Hill theorem thus also applies in the case we are dealing with a theory containing non-local terms in the action, where the gauge fields are not constrained to the plane while the fermions are, and this irrespective of the presence of the Fermi velocity $v_F<c$ which breaks explicitly the Lorentz invariance. Let us point out that Lorentz invariance can be broken in an even more severe way, namely rotational symmetry breaking, once we abandon the linear regime due to the honeycomb lattice structure. This has important consequences for the $\pi$ electron description in such regimes \cite{IorioPais,Iorio2017}. For completeness, we have also derived the tree-level photon propagator for this theory, taking into account the CS term. Interestingly, for the RQED case, the parameter $\theta$ in front of the CS term is not a mass, as for QED$_{3}$, but somehow acts as a dimensionless suppressing factor in the photon propagator (see \eqref{photon_prop_tree}). Moreover, we computed the exact value of $\theta$ in case the four-component Dirac fermions are massive for two different realizations of the mass term, both relevant for graphene studies, see \eqref{action_gamma3} and \eqref{action_haldane} which is valid for both Lorentz invariant  and non-invariant case.

Our observations pave the road to investigate deeper the interconnection between the CS photon term and Haldane fermion mass in the specific case of RQED. Any interaction term or fermion mass has a direct influence in the vector and axial current channels which, in the context of graphene physics, provide us with relevant observables for transport phenomena. A mapping between the two sectors of the theory would also allow us to investigate how the presence of external electromagnetic fields effectively manifest itself in the fermion sector. An important piece of information will be encoded in the $\theta$-sector of the photon propagator, which we expect to be quite sensitive to such background fields.  Numerical estimates for the influence of the CS term on the Haldane mass and/or $\gamma^3$-Dirac mass making use of Dyson--Schwinger equations, along the same lines as the QED$_3$ study of \cite{Hofmann:2010zy}, are currently being prepared and will be reported in forthcoming work, also paying due attention to the r\^{o}le of the Fermi velocity $v_F$.  Moreover, we hope also to come back to establishing the one-loop exactness of the topological photon term, at least in case of constant background electric and magnetic field, which are of phenomenological relevance as outlined in our text. These backgrounds can be rephrased in terms of a single space-time dependent background gauge field, which itself remains constant in momentum space, up to appropriate (derivatives of) $\delta$-functions. It should thence be possible to construct the most general transverse tensor basis, in presence of both aforementioned background fields and Fermi velocity $v_F$, and apply similar techniques as outlined here.

\section*{Acknowledgments}
We thank S.~Hern\'{a}ndez-Ortiz, A.~Raya and C.~Villavicencio for useful correspondence. A.~Mizher is a beneficiary of a postdoctoral grant of the Belgian Federal Science Policy (BELSPO) and receives partial support from FAPESP under scholarship number 2016/12705-7, while the work of P.~Pais is supported by a PDM grant of KU Leuven.

\bibliography{topphotonmass_biblio}{}
\bibliographystyle{apsrev4-1}

\end{document}